# Computational Thinking Development in AI Agent Creation: A Mixed-Methods Study

Yimeng Sun[1] , Haiyang Xin[1[0009-0005-8002-6538]], Qiannan Niu[1], Shuang Li[1[0009-0005-9556-6166]], Lingyun Huang[2[000-0002-7336-7079]] and Gaowei Chen[3[0000-0002-6847-4013]]

[1] CocoRobo LTD, Hong Kong, China, sunyimeng@cocorobo.cc, Tony@cocorobo.cc, niuqiannan@cocorobo.cc, lishuang@cocorobo.cc
[2] The Education University of Hong Kong, Hong Kong, China, lingyunhuang@eduhk.hk
[3] The University of Hong Kong, Hong Kong, China, gwchen@hku.hk

**Abstract.** This mixed-methods study examined computational thinking (CT) development among 93 pre-high school students in a five-day AI agent creation workshop using CocoFlow, a no-code platform. Integrating pre-post assessments, behavioral logs, and interviews, we investigated CT development and how initial CT levels shape learning trajectories. Results revealed significant improvements in abstract thinking ($d$ = 0.71) and algorithmic thinking ($d$ = 0.70). Hierarchical regression identified iterative testing engagement as a predictor of self-efficacy gains ($\beta$ = .20, $p$ = .050). Notably, students with moderate initial CT levels demonstrated substantially greater gains than both high-CT and low-CT peers, revealing an "Optimal Development Zone" effect ($\eta^2$ = .55). Qualitative analysis showed moderate-CT students exhibited adaptive expertise, while high-CT students risked over-engineering and low-CT students struggled with task decomposition. These findings challenge linear learning assumptions and provide evidence for differentiated scaffolding in CT education.

## 1 Introduction

As artificial intelligence becomes increasingly integrated across educational and societal domains, individuals require new cognitive competencies to effectively collaborate with AI systems. In this context, computational thinking (CT) has emerged as a critical 21st-century competency [1]. CT represents a systematic thinking process centered on problem formulation and translating solutions into algorithmic steps executable by computational agents [2-3]. This cognitive approach extends beyond programming, encompassing abstraction, decomposition, algorithmic design, pattern recognition, and debugging [4]. These processes develop problem-solving frameworks that transfer across diverse contexts.

Pedagogical approaches for cultivating CT continue to evolve beyond traditional programming instruction. No-code platforms represent a pedagogical shift by lowering technical barriers through visual interfaces and pre-built components, shifting cognitive demands from syntactic precision to higher-order design activities such as systems thinking, algorithmic logic, and iterative refinement [5]. Platforms enabling AI agent creation present particularly promising opportunities. Students design conversational agents responding to user inputs—activities requiring them to decompose ambiguous requirements, abstract business logic into decision models, design algorithmic sequences, and iteratively test and refine agent behaviors. By creating AI systems rather than merely using them, students may develop a "creator's perspective" that deepens understanding of human-AI collaboration.

Despite this potential, critical gaps remain. While emerging research shows AI-integrated instruction enhances students' systems thinking and problem decomposition [6-7], existing studies predominantly examine students learning with AI as users rather than creators. More importantly, studies capture CT at discrete time points rather than examining its dynamic developmental process. We lack understanding of how students' subjective self-perceptions (e.g., confidence in CT abilities) evolve alongside objective behavioral patterns (e.g., design and debugging strategies) throughout learning. Second, the role of individual differences requires further investigation. Learning theories, including Vygotsky's [8] Zone of Proximal Development and the Expertise Reversal Effect [9], suggest learners at different competency levels may exhibit distinct trajectories. Yet key questions remain: Do students with different initial CT levels show differential gains? How do their learning challenges and strategies differ?

This mixed-methods study addresses these gaps by examining CT development among 93 pre-high school students participating in a five-day AI agent creation workshop using a no-code platform. The study integrates quantitative pre-post assessments and behavioral logs with qualitative interviews to investigate two research questions:

- **RQ1: How do students develop CT competencies through AI agent creation?**
- **RQ2: How do individual differences in initial CT levels shape learning trajectories?**

By triangulating performance assessments, platform logs, and semi-structured interviews, this study provides new theoretical insights into CT development mechanisms in AI-integrated environments and offers empirical evidence for differentiated scaffolding in CT education.

## 2 Theoretical Framework

The study's theoretical framework integrates two complementary perspectives. First, the International Society for Technology in Education's (ISTE) operationalization of computational thinking provides a measurement framework, deconstructing CT into six interconnected elements: decomposition, pattern recognition, abstraction, algorithmic thinking, identification/analysis/implementation, and generalization/transfer [10]. This framework exhibits structural correspondence with the AI agent creation process, as students must decompose user requirements, abstract business logic, apply algorithmic thinking to design automated sequences, and refine agents through iterative testing.

Second, learning science theories explain individual differences. Vygotsky's [8] Zone of Proximal Development and Cognitive Load Theory [11] suggest optimal learning requires appropriately challenging tasks. The Expertise Reversal Effect [9] and Hatano and Inagaki's [12] distinction between Routine and Adaptive Expertise predict that students with different initial CT levels will follow distinct developmental trajectories.

# 3 Methods

## 3.1 Participants and Context

Ninety-three incoming high school freshmen (age M = 14.8, SD = 1.6; 54 males, 39 females) from a public high school in a major city in southern China participated in a five-day AI agent development workshop. Participants were recruited through voluntary enrollment without prior screening for programming experience. The workshop followed a progressive structure: Days 1-3 (9 hours) focused on AI theory and no-code platform training; Days 4-5 (12 hours) concentrated on autonomous project development, where students designed conversational AI agents in teams. Among the 93 participants, 32 volunteered for post-workshop interviews. All participants and guardians provided informed consent.

## 3.2 Data Sources

We employed a convergent mixed-methods design with pre-post assessments (Day 0 and Day 5), behavioral logging during autonomous project development (Days 4-5), and post-workshop interviews (n=32), integrating four data sources (Table 1).

**Table 1.** Data Sources and Analysis Methods

| Data Source | Type | Key Information | Analysis |
|---|---|---|---|
| CT Baseline Assessment | Objective | Bebras Challenge; initial CT level | K-means clustering ($k = 3$) |
| CT Self-Perception Scale | Self-report | 18 items, 3 dimensions; pre-post ($\alpha > .95$) | Paired t-tests; ANOVA |
| Platform Behavioral Logs | Objective | Usage patterns, iterations (Days 4-5) | Markov chains; Regression |
| Semi-Structured Interviews | Qualitative | 32 students, ISTE CT framework ($\kappa > .70$) | Thematic analysis |

**The No-Code AI Agent Platform.** The intervention employed CocoFlow, a web-based platform enabling students to create conversational AI agents without programming (Appendix). Students drag pre-built components from a module library to a visual canvas, constructing dialogue workflows that define how agents respond to user inputs. Key components include intent recognition modules (identifying user goals), entity extraction modules (capturing key information), and conditional logic nodes (if-then branches). The platform provides a real-time chat simulator for immediate testing and debugging. A pre-trained NLU engine handles intent recognition and entity extraction, allowing students to focus on system-level design. This design shifts cognitive demands from programming syntax to higher-order competencies: understanding AI capabilities, designing user-centered interactions, decomposing tasks, and iterative refinement.

### 3.3 Data Analysis

We employed a mixed-methods analytical strategy integrating quantitative and qualitative data. Quantitative analysis included exploratory factor analysis ($\alpha > .85$, KMO > .80), paired-samples t-tests for pre-post comparisons, K-means clustering ($k = 3$) for individual differences, Markov chain models for behavioral transitions, and hierarchical regression for self-efficacy predictors. Semi-structured interviews ($n = 32$) were independently coded by two researchers (κ > .70) to contextualize quantitative findings. All analyses used R and MAXQDA 24 ($\alpha = .05$).

## 4 Results and Findings

This section presents findings from our mixed-methods investigation of how students develop computational thinking competencies through AI agent creation, with particular attention to individual differences in learning trajectories.

### 4.1 RQ1: Development of CT Competencies

Students demonstrated significant CT development across three interconnected dimensions: self-perception, behavioral patterns, and learning processes.

**Self-Perception Improvements.** Paired-samples t-tests ($N = 84$) revealed significant post-intervention improvements across all three CT dimensions (Table 2). Abstract thinking ($d = 0.71$) and algorithmic thinking ($d = 0.70$) showed large effect sizes, while pattern recognition and generalization demonstrated a medium effect ($d = 0.35$). These findings indicate that AI agent creation significantly enhanced students' confidence in core CT competencies, particularly in abstraction and algorithmic reasoning. The substantial improvements in abstract and algorithmic thinking reflect students' intensive practice in translating user requirements into AI agent designs and configuring conditional logic. The relatively modest improvement in pattern recognition ($d = 0.35$) suggests this higher-order capability may require longer-term cultivation beyond a five-day intervention.

**Table 2.** Pre-Post Changes in CT Self-Perception (N=84)

| Dimension | Pre-test M (SD) | Post-test M (SD) | Mean Difference [95% CI] | $t(83)$ | $p$ | Cohen's d |
|---|---|---|---|---|---|---|
| Abstract Thinking (CT1) | 5.16 (0.99) | 5.68 (0.87) | 0.52 [0.36, 0.67] | 6.55 | < .001 | 0.71 |
| Algorithmic Thinking (CT2) | 5.98 (1.18) | 6.59 (0.99) | 0.61 [0.42, 0.80] | 6.46 | < .001 | 0.7 |

| Pattern Recognition & Generalization (CT3) | 5.66 (1.01) | 5.91 (0.88) | 0.25 [0.10, 0.41] | 3.2 | 0.002 | 0.35 |
|---|---|---|---|---|---|---|

**Behavioral Evolution.** Markov chain analysis of Days 4-5 platform logs revealed behavioral shifts toward goal-oriented optimization (Figure 1). Redundant design-implementation switching decreased (transition probabilities decreased by 0.12 and 0.08, $p < .001$), while optimization-related transitions increased (by 0.061 and 0.076, $p < .03$), indicating more efficient problem-solving and systematic refinement.

**Fig. 1.** Behavioral Transition Network - Day 4 to Day 5

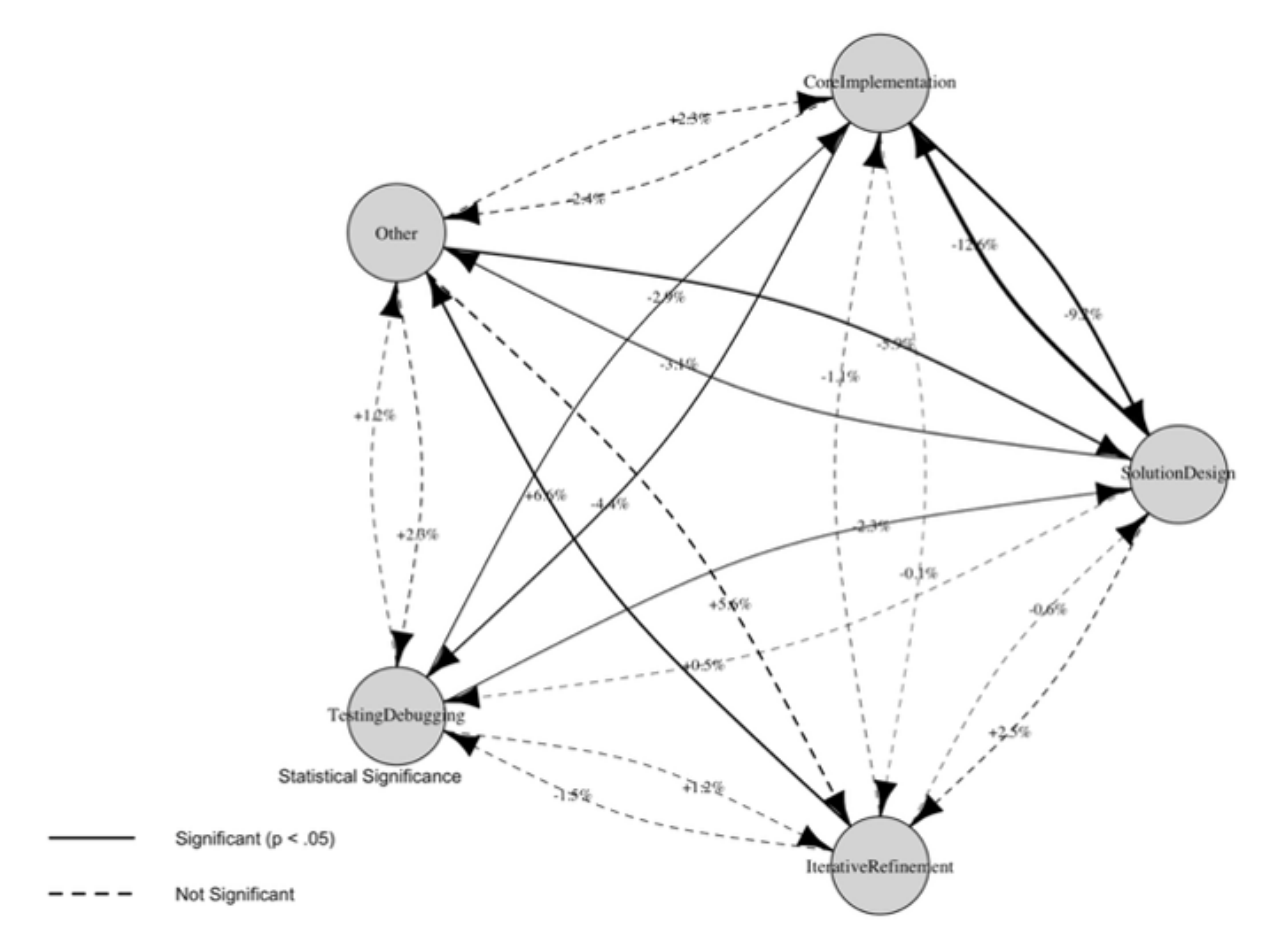

**Testing as a Key Mechanism.** Hierarchical regression identified iterative testing engagement as the sole behavioral predictor of CT self-efficacy gains (Table 3). After controlling for initial self-efficacy ($\beta = -.52, p < .001$), iterative testing positively predicted gains ($\beta = .20, p = .050, \Delta R^2 = .035$). Bootstrap validation (BCa 95% CI [1.10, 116.78]) and cross-validation ($R^2 = .38$) confirmed model stability.

**Table 3.** Hierarchical Regression Predicting Self-Efficacy Gains (N=80)

| Model & Variable | B | SE | $\beta$ | $t$ | $p$ |
|---|---|---|---|---|---|
| Model 1 (Baseline) | | | | | |
| (Constant) | 42.14 | 5.9 | | 7.14 | < .001 |
| Initial Self-Efficacy | -0.32 | 0.058 | -0.53 | -5.55 | < .001 |
| Model 2 (Final Model) | | | | | |
| (Constant) | 39.9 | 5.9 | | 6.76 | < .001 |
| Initial Self-Efficacy | -0.3 | 0.058 | -0.52 | -5.17 | < .001 |

| Iterative Testing Engagement | 57.83 | 29.02 | 0.2 | 1.99 | 0.05 |
|---|---|---|---|---|---|

*Note*. Model 1: $R^2$=.283; Model 2: $R^2$=.318, $\Delta R^2 = .035$, $F(1,77) = 3.97$, $p$=.050.

### 4.2 RQ2: Individual Differences in Learning Trajectories

K-means clustering ($k$=3) identified three learner profiles, revealing an "Optimal Development Zone" effect.

**The "Optimal Development Zone" Effect.** The three profiles showed markedly different characteristics: High-CT (n=37, minimal gains), Moderate-CT (n=21, substantial gains), and Low-CT (n=26, minimal gains). ANOVA revealed significant differences [ $F(2,84) = 51.64, p < .001, \eta^2 = .55$ ]. Moderate-CT students demonstrated the greatest gains (Table 4), significantly outperforming both High-CT ($\Delta = 17.62, p < .001$) and Low-CT groups ($\Delta = 18.88, p < .001$).

Critically, the Moderate-CT group demonstrated the greatest gains (Table 4), significantly outperforming both High-CT (mean difference=17.62, $p < .001$) and Low-CT groups (mean difference=18.88, $p < .001$). High-CT and Low-CT groups showed similar, modest gains (mean difference=-1.27, $p = .761$), despite vast differences in absolute ability levels.

**Table 4.** Self-Efficacy Gains by CT Level Group

| Group Comparison | Mean Difference | 95% Confidence Interval | $p$-value |
|---|---|---|---|
| Mid-CT Group vs. High-CT Group | 17.62 | [13.03, 22.20] | < .001*** |
| Low-CT Group vs. High-CT Group | -1.27 | [-5.55, 3.02] | 0.761 |
| Low-CT Group vs. Mid-CT Group | -18.88 | [-23.88, -13.89] | < .001*** |

Semi-structured interviews (n=32) corroborated these patterns. Moderate-CT students exhibited adaptive strategies: "First, I imagined the simplest possible version... and successfully saved about 6-7 components." High-CT students sometimes over-engineered: "maybe my thinking is too complicated... I spent the whole morning on it." Low-CT students relied on external guidance: "the teacher said this idea was too naïve." These patterns suggest moderate initial CT provides optimal balance between technical capability and cognitive flexibility.

## 5 Discussion

This study investigated computational thinking (CT) development through AI agent creation, revealing two principal findings: (1) iterative testing engagement predicts self-efficacy gains, and (2) students with moderate initial CT levels demonstrate the greatest developmental benefits—an "Optimal Development Zone" effect.

### 5.1 Iterative Refinement as a Key Learning Mechanism

Regression analysis revealed that test-refine-debug cycles significantly predicted self-efficacy gains ($\beta = .20, p = .050$). This finding challenges traditional programming pedagogy's emphasis on initial correctness. In AI agent creation, where responses are inherently variable, the capacity to iterate effectively proves more valuable than designing perfectly on the first attempt. This mechanism aligns with Bandura's [13] conceptualization of mastery experiences as the most potent source of self-efficacy: each successful iteration provides concrete evidence of growing competence. The no-code platform's immediate feedback loop transformed potential frustrations into productive learning opportunities.

### 5.2 The Optimal Development Zone: A Non-Linear Learning Trajectory

A key finding is that students with moderate initial CT demonstrated significantly greater self-efficacy improvements than both high-CT and low-CT peers, who showed similar modest gains despite vastly different starting points.

We interpret this through three frameworks. First, Vygotsky's [8] Zone of Proximal Development suggests moderate-CT students found tasks appropriately challenging, while high-CT students experienced insufficient challenge and low-CT students faced cognitive overload [11]. Second, the Expertise Reversal Effect [9] explains high-CT students' constrained gains: established schemas limited openness to novel strategies. Moderate-CT students exhibited adaptive expertise [12], balancing systematic approaches with flexibility.

### 5.3 Theoretical Contributions and Practical Implications

This study makes three theoretical contributions. (1) demonstrating no-code AI agent creation effectively develops CT ($d = 0.71$ for abstraction, $d = 0.70$ for algorithmic thinking); (2) identifying iterative testing as a behavioral predictor of gains; (3) providing evidence for non-linear developmental trajectories.

For educational practice, differentiated scaffolding is essential: moderate-CT students benefit from open-ended challenges; high-CT students need constraint-based tasks; low-CT students require structured pathways with agent modification tasks.

### 5.4 Limitation

Several limitations contextualize these findings. First, the five-day workshop represents a short-term intervention; longitudinal studies are needed to assess persistence and transfer. Second, due to time constraints, we were unable to administer post-intervention objective CT assessments. While the Bebras baseline test provided objective initial CT measures for grouping students, future research should include objective pre-post CT assessments to complement self-efficacy measures and provide convergent validity.

# 6 Conclusion

This study demonstrates that AI agent creation provides an effective context for developing computational thinking competencies, with effectiveness depending on students' initial readiness and engagement in iterative refinement processes. Two principal findings emerge. First, iterative testing engagement significantly predicts self-efficacy gains, indicating that the quality of students' interaction with AI systems matters more than initial design accuracy. Second, the Optimal Development Zone effect reveals a non-linear relationship between initial CT levels and learning outcomes, with moderate-CT students benefiting most while both high-CT and low-CT peers showed constrained gains, underscoring the necessity of differentiated scaffolding. As AI technologies become increasingly prevalent, cultivating students' capacity to iteratively design, test, and refine AI systems represents a critical educational priority.

# 7 Appendix

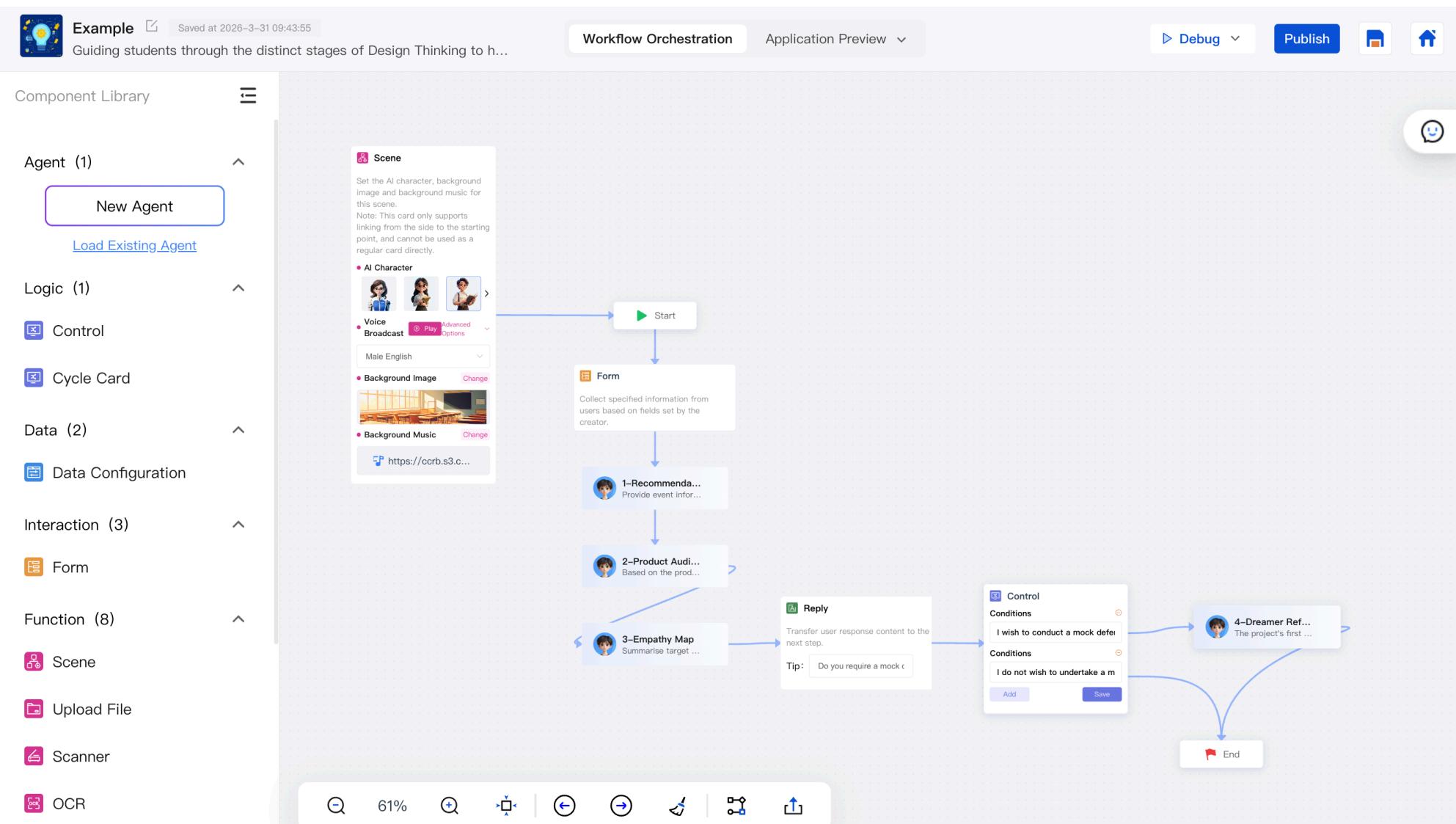